\documentclass[a4]{elsart5p}

\usepackage{amssymb,amsmath} 
\usepackage{bm,hyphenat,xspace} %,draftcopy}
\usepackage{graphicx,epsfig}
\usepackage{tabularx}
\usepackage{epsfig} %,pstricks,psfig,epic,eepic,pst-coil}
\newcommand{\email}[1]{\ead{#1}}

\newcommand {\mbf}[1]{{\mathbf{#1}}}

\newcommand {\mcu}{\mathcal{U}}

\newcommand{\A}[2]{{}^{#1}\mathrm{#2}}

\newcommand{\He}{{}^3\mathrm{He}}
\newcommand{\HE}{{}^4\mathrm{He}}
\newcommand{\Hh}{{}^3\mathrm{H}}
\newcommand{\Hd}{{}^2\mathrm{H}}

\begin{document}
%\linenumbers

\begin{frontmatter}

\title {Deuteron-${}^{3}\mathrm{He}$ scattering using
nucleon-${}^{3}\mathrm{He}$ optical potentials 
fitted to four-body amplitudes}

\author{A.~Deltuva},
\email{arnoldas.deltuva@tfai.vu.lt}
\author{D. Jur\v{c}iukonis}
\email{darius.jurciukonis@tfai.vu.lt}

%\affiliation{
\address{
Institute of Theoretical Physics and Astronomy, 
Vilnius University, Saul\.etekio al. 3, LT-10257 Vilnius, Lithuania}
%Centro de F\'{\i}sica Nuclear da Universidade de Lisboa, 
%P-1649-003 Lisboa, Portugal }

%\received{May 31, 2022}

%\pacs{24.10.-i, 21.45.-v, 25.45.Hi, 25.40.Hs}

\begin{abstract}
Deuteron-${}^{3}\mathrm{He}$ reactions in the 15 to 40 MeV range are studied using a three-body model
where the constructed nonlocal optical potentials rely on rigorous nucleon-${}^{3}\mathrm{He}$  scattering calculations.
The differential cross section for the elastic scattering and neutron transfer reaction 
is predicted quite well up to 90 deg scattering angles. The importance of the Pauli term in complex potentials
is demonstrated. 
\end{abstract}


\end{frontmatter}
% \maketitle

% -----------------

\section{Introduction \label{sec:intro}}

{%\bf 
Introduction  of optical potentials into the nuclear reaction theory
 enabled the reduction of a many-nucleon problem,  encountered in the
nucleon-nucleus scattering,
to an effective two-body problem.}
The enormous complexity of the many-body problem in the continuum for a long time restricted 
the construction  of optical potentials to the phenomenological approach, where
model parameters were adjusted to the experimental data, such as
in the Chappel Hill \cite{CH89},
Koning and Delaroche \cite{koning}, Weppner et al. \cite{weppner:op} and many other optical potentials.
Many-nucleon structure calculations using various methods
such as the microscopic mean field, the no-core shell model, Green's function Monte Carlo, coupled cluster approach,
self-consistent Green's function progressed significantly in the last decades, opening the doors
for approximate extensions to the continuum and  microscopic calculations of the optical potentials
as reviewed in Ref.~\cite{microOP:23}. For example,
quite a typical approach is folding the microscopically calculated
nuclear densities with the nucleon-nucleon interaction  \cite{gennari:18a,vorabbi:22a,furumoto:mgop},
 either bare or $G$-matrix one.
A prerequisite for this approach, limited to the first order scattering term, is high enough beam energy.
This way some methods, like those based on the mean-field description, were able to develop
microscopic optical potentials also for heavy nuclei above mass number $A=200$.
In contrast, the lightest nuclei such as $\Hh$ and $\He$
remain beyond the reach of those methods as the individual character
of their constituent nucleons plays an important role, and it is not even clear to what extent
the optical potential description can be successful. 
{%\bf 
Since nucleon-$\HE$  scattering calculations are  available
only below the breakup threshold \cite{nollett:07a,quaglioni:08a,lazauskas:18a},
we study the lighter isotope $\He$, with rigorous nucleon-$\He$ scattering calculations available at energies
well above the breakup threshold \cite{deltuva:13c,deltuva:14b}. 
Furthermore, $\He$ has  the nucleon separation energy  around 5 MeV, which is quite a typical excitation energy for many light
nuclei, in contrast to exceptionally tightly bound $\HE$, and thus might be more suitable to draw conclusions.
}
Therefore
we aim to construct the optical models for proton ($p$) and neutron ($n$) interactions
with $\He$ based on rigorous continuum calculations, a unique feature among the optical potentials.
Beside the quality in reproducing the experimental { nucleon-nucleus} scattering data, the further criterion
is the ability to describe more complicated reactions such as the deuteron-nucleus scattering.
Thus, another goal of the present work is the application of the developed nucleon-$\He$ potentials to the
deuteron-$\He$ elastic scattering, breakup, and the neutron transfer reaction $\He(d,p)\HE$
at energies well above the breakup threshold of the involved nuclei.
As the transfer reaction at very low energy is of high importance for the termonuclear fusion,
in the regime below the breakup threshold it has also been studied using the no-core shell model
with continuum \cite{navratil:12a,hupin:19a}.

A somehow similar idea has been explored in the past for $\A{5}{H}$ resonance study in the
three-body $\Hh+n+n$ model by developing effective neutron-$\Hh$ potential
\cite{diego:tnn}.
It contained several local Gaussian terms whose
parameters were fitted to phase shifts  below the inelastic threshold. Consequently, the imaginary
part was  vanishing. An important feature of that potential was a strong partial-wave dependence
of its parameters, quite common in few-nucleon systems. In the present work we aim to develop
complex potentials  valid over a broader energy range with open inelastic channels,
but we also expect the need for strongly partial-wave dependent parameters.

Section \ref{sec.nh} recalls microscopic four-nucleon reaction calculations,
section \ref{sec.op} describes and validates the developed nucleon-$\He$ optical potentials, while
section \ref{sec:eq} reminds the three-cluster deuteron-$\He$  scattering equations.
Sections \ref{sec:res} and \ref{sec:con} contain the deuteron-$\He$  scattering results
and our conclusions, respectively.
%Note that we use natural units $\hbar=c=1$.

\section{Four-body calculation of the nucleon-$\He$ scattering \label{sec.nh}}

A rigorous quantum-mechanical description for the 
nucleon scattering from the three-nucleon bound state can be given by the
Alt, Grassberger and Sandhas (AGS) equations 
\cite{grassberger:67} for transition operators $\mcu_{\beta \alpha}$
that constitute a momentum-space integral equation
formulation of the Faddeev-Yakubovsky (FY) four-particle theory \cite{yakubovsky:67}.
Previous benchmark calculations \cite{viviani:11a,viviani:17a}
with alternative theoretical frameworks, 
namely, the hyperspherical harmonics  expansion method
\cite{viviani:01a,kievsky:08a} and the coordinate-space FY equations 
\cite{lazauskas:04a,lazauskas:09a}, performed below the breakup threshold, revealed
good agreement between the three methods, confirming their reliability.
The symmetrized form of the AGS equations \cite{deltuva:07a}, most convenient
for the four-nucleon system in the isospin formalism, reads
\begin{subequations}  \label{eq:AGS}   
\begin{align}  
\mcu_{11}  = {}&  -(G_0 \, t \, G_0)^{-1}  P_{34} -
P_{34} U_1 G_0 \, t \, G_0 \, \mcu_{11}  \nonumber \\ 
{}& + U_2   G_0 \, t \, G_0 \, \mcu_{21}, \label{eq:U11}  \\
\label{eq:U21}
\mcu_{21} = {}&  (G_0 \, t \, G_0)^{-1}  (1 - P_{34})
+ (1 - P_{34}) U_1 G_0 \, t \, G_0 \, \mcu_{11},
\end{align}
\end{subequations}
where $G_0$ is the free resolvent that gives rise to energy-dependence of the transition
operators, $t = v + v G_0 t$ is the two-nucleon transition
operator derived from the two-nucleon potential $v$ including the screened
Coulomb force for the two-proton pair, and
\begin{gather} \label{eq:AGSsub}
U_\alpha =  P_\alpha G_0^{-1} + P_\alpha t\, G_0 \, U_\alpha
\end{gather}
are subsystem transition operators. The subscripts 1 and 2 label the
(12,3)4 and (12)(34) partitions, while $P_\alpha$ and $P_{34}$ are permutation
operators, explained in Ref.~\cite{deltuva:07a} together with other details.
The on-shell elements of $\mcu_{11}$ between the neutron-$\He$ or proton-$\He$
channel states yield the respective amplitudes for the elastic scattering.

The AGS equations (\ref{eq:AGS}) are solved in the momentum-space partial-wave representation
where they become a large system of up to 30000 equations in three continuous variables,
the Jakobi momenta. Two special procedures are employed:
(i) screening and renormalization method \cite{taylor:74a,semon:75a,alt:80a,deltuva:05a}
to include the Coulomb interaction between the protons, and
(ii) the complex-energy method with special weights \cite{deltuva:12c} to deal with integrable but highly 
complicated singularities in the integral equation kernel; see
Refs.\cite{deltuva:13c,deltuva:14b,deltuva:12c}
for more details and example results for scattering observables.

\section{Nucleon-$\He$ optical potential \label{sec.op}}

%The calculation of nucleon-$\He$ scattering amplitudes is briefly described in previous section
%and presented in Refs.\cite{deltuva:13c,deltuva:14b}.
Typically, the parameters of phenomenological optical potentials are determined by fitting
the experimental data for rather few selected scattering observables, such as the differential cross section,
analyzing power and inelastic cross sections.
As we base our nucleon-$\He$ optical potential on rigorous microscopic calculations,
the quantities to be reproduced are the theoretical elastic scattering amplitudes. Furthermore,
since the transition operators
contain full information on the physical system including inelastic processes,
reproducing elastic amplitudes implies also reproducing the predictions for the inelastic cross section,
that can be calculated via optical theorem from the imaginary part of the forward scattering amplitude.
In the neutron-$\He$ case it has contributions not only from the breakup but also from the charge exchange
$\He(n,p)\Hh$ and proton transfer $\He(n,d)\Hd$ reactions.
One might perhaps question whether those channels can be accounted for by the optical potential,
since the proton-$\Hh$ threshold is even lower in energy than the neutron-$\He$ one.
However, this is relevant near the threshold only, while in the considered energy regime 
the elastic neutron-$\He$ cross section exceeds the ones for charge exchange and proton transfer by one
order of magnitude \cite{deltuva:14b}.  

Previous studies \cite{deltuva:13c,deltuva:14b} found that the 
experimental data for nucleon-$\He$ scattering in the 10 to 35 MeV regime are best reproduced
using the inside-nonlocal outside-Yukawa (INOY04) two-nucleon 
potential  by Doleschall \cite{doleschall:04a,lazauskas:04a}. Thus, the scattering amplitudes
obtained solving four-nucleon AGS equations with this potential will be used as input
for the construction of  nucleon-$\He$ optical potentials.

We demand that solutions of the two-body Lippman-Schwinger equation
\begin{equation}  \label{eq:T}
  T_N = V_N + V_N G_0 T_N
\end{equation}
with the nucleon-$\He$ optical potentials $V_N$ reproduce accurately the
respective microscopic elastic scattering amplitudes over the energy range of roughly 13 to 30 MeV, which
is relevant for the application to the deuteron-$\He$ scattering and 
where the four-body calculations are available. 
There is some arbitrariness in choosing the form of the potential. Optical potentials in general are energy-dependent,
however, it is well known that the energy dependence is weaker if the potential is nonlocal in the
coordinate space.
Using  the proton-$\A{24}{Mg}$ elastic and inelastic scattering as example \cite{deltuva:23b} we demonstrated
recently that an energy-independent nonlocal optical potential may provide a reasonable description of the
experimental data in finite energy range. For this reason we assume  the nonlocal form also for 
 the nucleon-$\He$ optical potential, i.e.,
 \begin{equation}  \label{eq:Vdj}
V_N(\mbf{r}',\mbf{r}) =  \frac12 \big[ H(|\mbf{r}'-\mbf{r}|) V(r) + V(r')H(|\mbf{r}'-\mbf{r}|) \big],
\end{equation}
 where $\mbf{r}'$ and $\mbf{r}$ are final and initial distances between particles, and 
  \begin{equation}  \label{eq:Hx}
 H({x}) = \pi^{-3/2} \rho^{-3} e^{-(x/\rho)^2}
  \end{equation}
%  \end{subequations}
  is the nonlocality function with the nonlocality range $\rho$.
  As in Ref.~\cite{deltuva:23b} we use typical value $\rho=1$ fm.
 The local part we represent by several Gaussian terms as it is quite common for effective potentials
 \cite{furumoto:mgop,diego:tnn}, i.e.,
 \begin{equation}  \label{eq:Vy}
   \begin{split}
     V(r) = {}& \sum_{j=1}^2 V_j e^{-(r/R_j)^2}  + i W_c e^{-(r/R_w)^2} \\ {}&
     + [V_{s} e^{-(r/R_s)^2} + iW_{s} e^{-(r/R_{w})^2}]
       ( \delta_s \mbf{s}_N\cdot \mbf{L} + \delta_t \hat{S}_{12})
     %  + V_t e^{-(r/R_t)^2} S_{12}
   \end{split}
 \end{equation}
 where $V_k$ and $W_k$ are strengths of various real and imaginary terms,  $R_k$ are their Gaussian radii,
 and  $\mbf{s}_N$, $\mbf{L}$ and $\hat{S}_{12}$ are the nucleon spin, nucleon-$\He$ orbital angular momentum
 and tensor \cite{stoks:94a} operators, respectively. Additionally, $\delta_s$ and $\delta_t$ being either 1 or 0
 control the presence/absence of spin-orbit and tensor terms.
 The proton-$\He$ potential is supplemented by the Coulomb force, which below the Coulomb radius of 1.7 fm
 is taken as the potential of uniformly charged sphere.
 First fitting attempt using single parameter set for all partial waves was not successful, consistently
 with previos neutron-$\Hh$ studies \cite{diego:tnn}, and
 confirming our expectation for a need of partial-wave dependent parameters.
 We therefore fitted each partial wave separately. In doing this we tried to keep as few fitting parameters
 as possible. That is, in many cases some terms in Eq.~(\ref{eq:Vy}) could be set to zero, or at least
 share the same radius. Furthermore, we found that the tensor term $\delta_t=1$ is only important for
 coupled partial waves with $L= J \pm 1$, $S=1$, while the spin-orbit term $\delta_s=1$ is only important for
 coupled partial waves with $L= J$, $S=0,\, 1$, where $J$ and  $S$ is the total angular momentum and spin, respectively.
Thus, tensor and spin orbit terms are not included simultaneously, and are missing in the uncoupled waves.
 We also found that in the regime up to 30 MeV the partial waves with $L>3$ can be safely neglected.
In order to estimate uncertainties, we developed several parameter sets for optical potentials. Quite typically,
they differ in range and strength, i.e., smaller radii imply larger strengths. While some changes can be seen in small components
of two-body scattering amplitudes on a fine scale, we verified  that the  predictions for the nucleon-$\He$ and deuteron-$\He$
 scattering observables are not visibly affected.

% The resulting optical potential parameters are collected in Table \ref{tab:1}.

 Except for the ${}^1S_0$ neutron-$\He$ partial wave all the other nucleon-$\He$ 
 $L=0$ waves are Pauli repulsive.
 This is reflected by mostly positive $V_j$ values in those waves.
 Another way \cite{kukulin:84a} to take into account this Pauli repulsion is to use attractive potential
supplemented by a strong repulsive
nonlocal term  $ |b\rangle \Gamma \langle b|$. % with large positive  $\Gamma$
This approach is often used in simple nuclear structure models
for the nucleon-nucleus real binding potential to project out the state $|b\rangle$
corresponding to an occupied shell. An example close to our present study is the real low-energy
nucleon-$\HE$ potential in the $L=0$ state \cite{thompson:00} .
While the differences in the two approaches for three-body bound state calculations
 are moderate at most, in 
 scattering calculations they become tremendous \cite{thompson:00,deltuva:06b}.
 To the best of our knowledge, such a Pauli-repulsive term has not yet been included into complex
 optical potentials, and will be investigated in the present work.
% The state $|b\rangle$ should simulate the nucleon state in $\He$ that can not accomodate
% one more $S$-wave nucleon. %Since both deuteron and  $\He$ are predominantly $S$-wave states,
 The $S$-wave component   of the deuteron and  $\He$
 overlap $\langle d q (L=0) |\He \rangle = \langle q |b \rangle $, $q$ being the spectator nucleon momentum,
is an appropriate representation of 
{ the occupied $L=0$ state in $\He$.}
For simplicity we add the same term $ |b\rangle \Gamma \langle b|$ to the
potential (\ref{eq:Vdj}) in the ${}^3S_1$ wave and, for proton-$\He$, in the ${}^1S_0$ wave,
and refit the parameters of (\ref{eq:Vy}). % were refitted and are collected in Table \ref{tab:2}.
The strength of the Pauli-repulsive term is set to $\Gamma=1$ GeV, but we verified that results become largely independent
of $\Gamma$ once it exceeds few hundred MeV.

Furthermore,
a modification of the  neutron-$\He$ potential in the ${}^1S_0$ partial wave is needed for the calculation
of the $\He(d,p)\HE$ reaction. The potential must be real to simulate the $\HE$ nucleus as the
bound state of $\He$ and neutron, though due to its large binding energy of about 20.6 MeV this is not a
good model. We demand the effective neutron-$\He$ potential to reproduce not only this binding energy,
but also that the bound-state wave function mimics (up to a factor) 
the $\langle \He | \HE \rangle$ overlap function, ensuring that the spatial and momentum
distribution of the neutron in our model is similar to that in the   $\HE$ nucleus.
{%\bf
  To estimate the uncertainties, we used two choices for this binding potential (\ref{eq:Vy}), namely,
$V_1 = -2V_2 = -132.848$ MeV and $V_1 = -V_2 = -156.113$ MeV, with radii $R_1=2R_2=1.7$ fm in both cases.
Such approach of approximating the single-particle wave function by an overlap becomes quite
common in including the many-body nuclear structure information into the few-body description of nuclear reactions
\cite{deltuva:24a}. 
 In Faddeev or AGS few-body calculations the bound-state wave function must be normalized to unity, thus,
the calculated single-particle theoretical cross section has to be  multiplied by the corresponding norm of
the overlap, i.e., the  spectroscopic factor (SF), as explained in detail in Ref.~\cite{deltuva:24a}.
The SF from our microscopic $\He$ and $\HE$  calculations  with INOY04 potential equals to 1.65.
A similar approach in representing the overlaps  by solutions of the single-particle Schr\"odinger equation
was developed in Ref.~\cite{brida11}. Despite using different nuclear Hamiltonian, the resulting overlap
and SF $\approx 1.6$ turn out to be close to ours. The agreement becomes even better in a
 recent update \cite{Bob-overlaps} using soft nuclear forces, 
leading to SF $=1.65$.}

\begin{figure}[!]
\begin{center}
\includegraphics[scale=0.64]{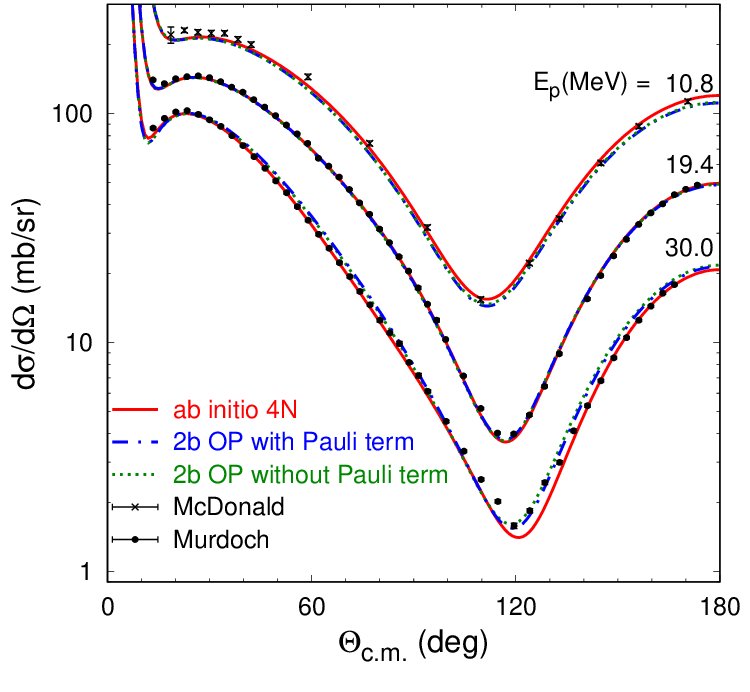}
\end{center}
\caption{\label{fig.ps}
  Differential cross section for the proton-$\He$ elastic scattering 
  at 10.8, 19.4 and  30 MeV beam energy as function of the c.m. scattering angle.
  Results of microscopic four-nucleon calculations (solid curves) are
  compared with predictions obtained using two-body optical potentials (2b OP),
  either without (dotted curves) or with  (dashed-dotted curves) the Pauli term.
The experimental data are from Refs.~\cite{mcdonald:64,murdoch:84a}.
}
\end{figure}

\begin{figure}[!]
\begin{center}
\includegraphics[scale=0.64]{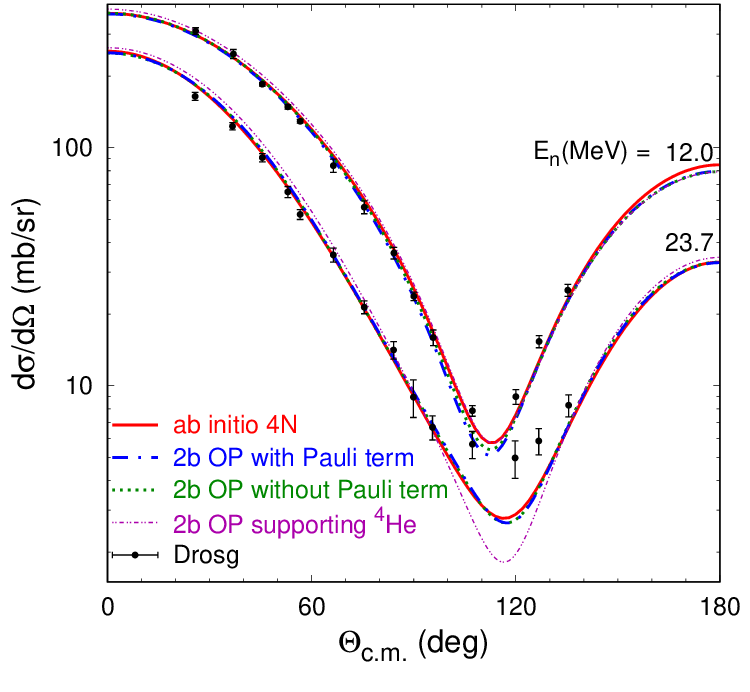}
\end{center}
\caption{\label{fig.ns}
  Differential cross section for the neutron-$\He$ elastic scattering 
  at 12 and  23.7 MeV beam energy. Curves as in Fig.~\ref{fig.ps}, while
  additional dashed-double-dotted curves label results with a real
  neutron-$\He$ potential in the ${}^1S_0$ partial wave supporting the $\HE$ bound state.
The experimental data are from Ref.~\cite{drosg:74a}.
}
\end{figure}

 We demonstrate the quality of our optical potentials by comparing with results of 
 microscopic four-body calculations. As examples in Figs.~\ref{fig.ps}  and \ref{fig.ns} we show the 
 proton-$\He$ and neutron-$\He$ differential cross sections, 
 achieving quite a satisfactory agreement between the predictions of two- and four-body models.
The agreement is almost perfect around the middle of the considered
 energy region, but some deviations occur at lowest and highest energies. However, even those deviations
 are smaller than the spread of the four-nucleon calculations obtained with different
 realistic two-nucleon potentials \cite{deltuva:13c,deltuva:14b}.
 The only more sizable deviation occurs near the neutron-$\He$  differential cross section minimum
 at higher energy when using a real potential in the ${}^1S_0$ partial wave supporting the $\HE$ bound state
 and thereby not fitted to the scattering amplitudes.
A further evidence of the accuracy can be found in the Supplemental material, together 
with the parameters of the developed optical  potentials.

\section{Three-body AGS equations \label{sec:eq}}

Deuteron scattering from a nucleus $A$ has been calculated in many works using
the Faddeev or its equivalent AGS transition operator formalism,
see Ref.~\cite{deltuva:06b} for the deuteron-$\HE$ example.
The three-body transition operators are obtained from the AGS equation
\begin{equation}  \label{eq:Uba}
U_{ba}  = \bar{\delta}_{ba} \, G_{0}^{-1}  +
\sum_{m=p,n,A}    \bar{\delta}_{bm} \, T_{m}  \,
G_{0} U_{ma},
\end{equation}
where the usual odd-man-out notation is used,
$\bar{\delta}_{ba} = 1 - \delta_{ba}$, the nucleon-nucleus transition
operators $T_p$ and $T_n$ are given by Eq.~(\ref{eq:T}), and
$T_A$ by an analogous equation with the neutron-proton potential.
For consistency we take the INOY04, though we verified that using other realistic potentials
yields very similar results.
The solution of the scattering  equations (\ref{eq:Uba}) is again performed  using the momentum-space 
partial-wave representation, while more technical details can be 
found in  Refs.~\cite{deltuva:06b,deltuva:23b} and references therein.

On-shell matrix elements
of $U_{ba}$ taken between the two- or three-cluster  channel states
determine the physical transition amplitudes for the respective reactions \cite{deltuva:06b},
from where the differential cross sections are calculated,
as for example outlined in  Ref.~\cite{deltuva:14b}.

\section{Results \label{sec:res}}

\begin{figure*}[!]
\begin{center}
\includegraphics[scale=0.8]{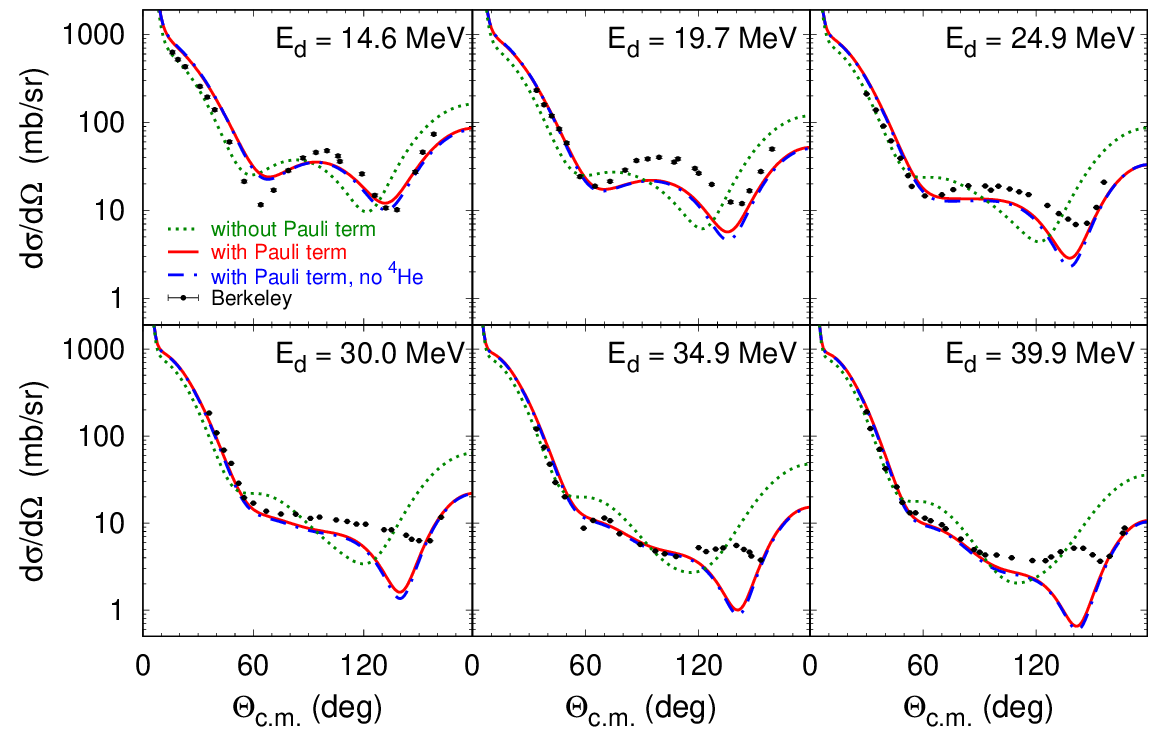}
\end{center}
\caption{\label{fig:del}
Differential cross section for deuteron-$\He$ elastic scattering in the energy range 14.6 to 39.9 MeV.
Predictions using optical potentials with/without Pauli term and with/without bound neutron-$\He$
state are compared with the experimental data from Ref.~\cite{d3he1540}.
}
\end{figure*}

Using the nonlocal interaction models described in Sec.~\ref{sec.op} we solve
the AGS equations (\ref{eq:Uba}) and calculate deuteron-$\He$ differential cross sections
at deuteron beam energy $E_d = 14.6$, 19.7, 24.9, 30.0, 34.9 and 39.9 MeV, where the
experimental data from Berkeley laboratory \cite{d3he1540} are available.
In Fig.~\ref{fig:del} we compare experimental data for elastic differential cross sections
with three calculations, that isolate two dynamics ingredients:
(i) including or excluding the Pauli term, and (ii) including or excluding the
$\HE$ bound state by using real or complex  neutron-$\He$ potential in the  ${}^1S_0$ partial wave.
The latter effect of the bound state turns out to be very small, discernible only near the minimum
at large scattering angles, with no any visible sensitivity to the parametrization of the
binding potential. In contrast, the effect of the Pauli term is very large, significantly changing
the shape of the angular distribution. Though none of the predictions are in a perfect
agreement with data, those including the Pauli term are considerably closer to the data, especially in the shape
at intermediate and backward angles.
One may perhaps argue that  $E_d = 14.6$ MeV is too low for 
our optical potentials fitted from 13 to 30 MeV nucleon energy, as typically one takes 
energy-dependent nucleon optical potentials at $E_d/2$. This might explain a abit larger discrepancy at small angles.
On the other hand, one can question also the experimental data which does not show monotonic variation in energy,
for example, the 30 MeV data at forward angles lies between the 19.7 and 24.9 MeV data.
In contrast, our theoretical predictions vary smoothly with energy.

\begin{figure*}[!]
\begin{center}
\includegraphics[scale=0.8]{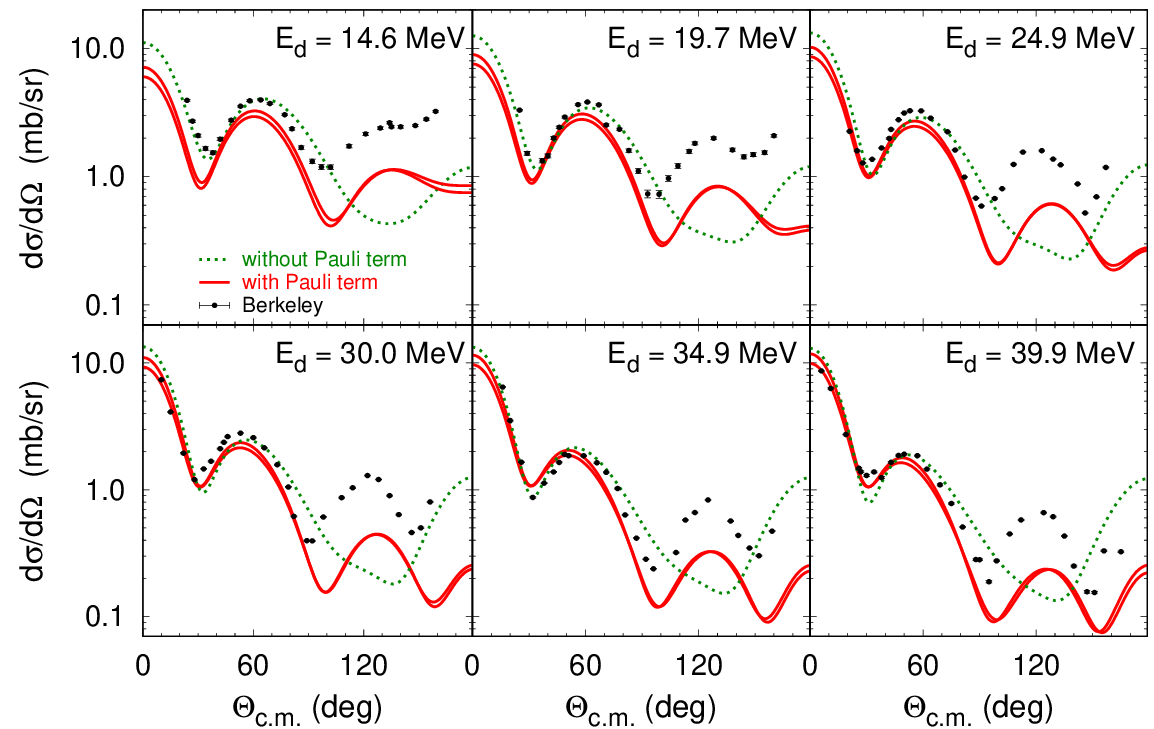}
\end{center}
\caption{\label{fig:dtr}
Differential cross section for the $\He(d,p)\HE$ reaction in the energy range 14.6 to 39.9 MeV.
Predictions using optical potentials with/without Pauli term
are compared with the experimental data from Ref.~\cite{d3he1540}.
The two solid curves correspond to different neutron-$\He$ binding potentials in the  ${}^1S_0$ partial wave.
}
\end{figure*}

In Fig.~\ref{fig:dtr} we compare experimental data for the transfer reaction $\He(d,p)\HE$
to several theoretical predictions including or excluding the Pauli term.
Again, the effect of this term turns out to be sizable, especially at larger scattering angles,
where the shape of the angular distribution is changed dramatically, leading to one more local
maximum of the differential cross section. Although the absolute value is significantly
underpredicted in this region, this change of shape due to the Pauli term is clearly supported by the experimental
data. The comparison of results with different neutron-$\He$ binding potentials in the  ${}^1S_0$ partial wave
reveals some sensitivity, but is far less important than the Pauli term.

\begin{figure}[!]
\begin{center}
\includegraphics[scale=0.8]{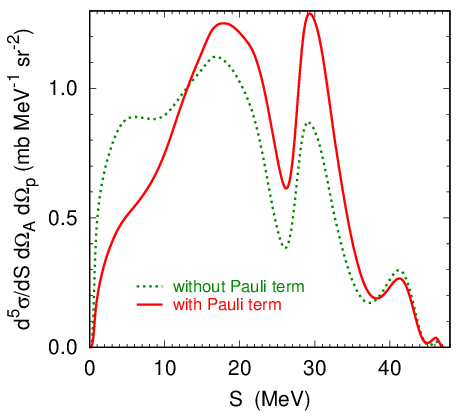}
\end{center}
\caption{\label{fig:dbr}
  Fivefold differential cross section for the deuteron breakup in collision with $\He$
  at 30 MeV beam energy as function of the arclength $S$.
  The final state kinematical configuration is characterized by $\He$ and proton scattering
  angles $(\theta_A,\phi_A)=(15^\circ,0^\circ)$ and $(\theta_p,\phi_p)=(45^\circ,180^\circ)$.
  Predictions of optical potentials with and without Pauli term
  are compared. The $\HE$ bound state is not supported.
}
\end{figure}

Finally, in Fig.~\ref{fig:dbr} we show that the Pauli term affects significantly also the
differential cross section in the
deuteron breakup reaction. In the case of fully exclusive breakup the observables are often shown
for fixed solid scattering angles $\Omega_a$ (polar $\theta_a$ and azimuthal $\phi_a$)
of two detected particles 
as functions of the arclength $S$ in the plane of their kinetic energies \cite{deltuva:06b}.
In Fig.~\ref{fig:dbr} those two particles are assumed to be $\He$ and proton.

\section{Conclusions \label{sec:con}}

We considered elastic, transfer and breakup reactions in  deuteron collisions with $\He$ nuclei.
We have not solved the underlying five-nucleon problem rigorously, but our study
does not rely on  the experimental information beyond the one contained in the realistic two-nucleon potential
INOY04. We used exact solutions of four-nucleon AGS scattering equations for the transitions operators
and, as an intermediate step,  developed nucleon-$\He$ optical potentials quite accuractely  reproducing
scattering amplitudes from  four-body calculations. We constructed also a model containing the Pauli term,
not included up to now into complex optical potentials.
Inserting those potentials into three-body AGS equations, that treat the $\He$ nucleus as an inert particle
but exatly account for the breakup of the deuteron, led  to predictions for deuteron-$\He$ reactions.

In cases of the elastic scattering and $\He(d,p)\HE$ reaction the comparison with the experimental 
differential cross section
revealed quite a good agreement up to about 90 deg scattering angles, but discrepancies remained at larger angles.

The model
including the Pauli term is considerably closer to the data, especially in the shape
at intermediate and backward angles. 
Noteworthy, our description of the $\He(d,p)\HE$ transfer  reaction is considerably more successful than
the DWBA analysis of Ref.~\cite{d3he1540}, though it used initial- and final-channel optical potentials
well fitted to the elastic deuteron-$\He$ and proton-$\HE$ data.

We speculate that the large-angle discrepancy, especially in the deuteron elastic scattering,
 is a signature for the reaction mechanism specific to 
very light nuclei but not included in our model. Namely, the reaction can proceed via the one-proton 
exchange, i.e., the $\He(d,\He)d$ reaction, where the deuteron picks one proton from $\He$ becoming a "new" 
$\He$, while the target $\He$ after loosing one proton becomes a "new" deuteron. 
Qualitatively the same reaction mechanism is present in the nucleon-deuteron scattering, where it is responsible 
for the backward angle cross section increase. In four-nucleon reactions it is a two-nucleon transfer that produces
a similar effect.

Our finding of general importance for nuclear reaction description is the significance of the Pauli term not
only for real but also for complex optical potentials. Further studies have to be performed
to evaluate the relevance of the Pauli term in heavier systems.

\vspace{1mm}

Authors thank P.~U.~Sauer for discussions during his stay at Vilnius University.
This work has received funding from the 
Research Council of Lithuania (LMTLT) under Contract No.~S-MIP-22-72.
Part of the computations were performed using the infrastructure of
the Lithuanian Particle Physics Consortium.

%%%%%%%%%%%%%%%%%%%%%%%%%%%%%%%%%%%%%%%%%%%%%%%%%%%%%%%%%%%%%%%%%%%%%%%%%%%%%

%\bibliographystyle{plbsty} 
%\bibliography{abbrev,pre80,80-89,90-99,200x,ad,book,clmb,exp,nreact,nstruct,4N,4Nex,dalpha} \end{document}

\end{document}

%\linenumbers

\begin{frontmatter}

\title {Supplemental material for "Deuteron-${}^{3}\mathrm{He}$ scattering using
nucleon-${}^{3}\mathrm{He}$ optical potentials fitted to four-body amplitudes"
}

\author{A.~Deltuva},
\author{D. Jur\v{c}iukonis}

%\affiliation{
\address{
Institute of Theoretical Physics and Astronomy, 
Vilnius University, Saul\.etekio al. 3, LT-10257 Vilnius, Lithuania}
%Centro de F\'{\i}sica Nuclear da Universidade de Lisboa, 
%P-1649-003 Lisboa, Portugal }

%\received{May 31, 2022}

%\pacs{24.10.-i, 21.45.-v, 25.45.Hi, 25.40.Hs}

\end{frontmatter}
% \maketitle

% -----------------

\section{Parameters of nucleon-$\He$ optical potentials \label{sec.op}}

\begin{table*} [h]
  \caption{Parameters of the proton-$\He$ optical potential. Strengths are in units of MeV, 
while radii are in units of fm. The parameters for spin-orbit or tensor forces are identical
in both coupled waves, and are listed just once.
\label{tab:1}
  }
%\begin{ruledtabular}
\setlength{\tabcolsep}{5pt}
%\begin{tabularx*}{\columnwidth}{l|*{9}{r}}
\begin{tabular}{l|*{9}{r}}
  \hline
  %                 1       2       9     10      7       8        5        6       13
  % arba            3       4      11     12      7*14                            
  ${}^{2S+1}L_J$ & $V_1$ & $R_1$ & $V_2$ & $R_2$ & $W_c$ & $R_w$ & $V_s$ & $R_s$  & $W_s$  
  \\ \hline
${}^1S_0$ &  82.032 & 2.186 & 9.031 & 3.977 & -2.193 & 2.075 &  &  &  \\
${}^3P_0$ & -46.343 & 1.840 & -4.510 & 3.347 & -1.917 & 1.679 &  &  &  \\
${}^3S_1$ & 100.000 & 2.022 & 5.202 & 4.277 & -7.000  & 1.850 & 1.813 & 2.653 & -1.105 \\
${}^3D_1$ & -70.000 & 1.344 & 2.328 & 3.919 & -21.000 & 1.850 &  &  &  \\
${}^1P_1$ & -19.957 & 2.116 & -3.841 & 3.841 & -9.517 & 1.125 & 39.364 & 1.332 & -1.000 \\
${}^3P_1$ & -14.354 & 2.116 & -9.416 & 3.841 & -8.426 & 1.125 &  &  &  \\
${}^3P_2$ & -95.729 & 1.613 & -6.754 & 3.391 & -2.589 & 1.297 & -5.735 & 2.411 & 0.865 \\
${}^3F_2$ &         &       & -5.184 & 3.391 & -5.741 & 2.714   &  &  &  \\
${}^1D_2$ & -16.981 & 2.367 & 4.304 & 3.902 & -5.000 & 2.004 & 10.000 & 1.782 & -0.763 \\
${}^3D_2$ & -20.000 & 2.367 & 2.731 & 3.902 & -10.024 & 2.004 &  &  &  \\
${}^3D_3$ & -60.000 & 1.875 & 2.563 & 3.872 & -5.000 & 2.228 &  &  &  \\
${}^1F_3$ &         &       & -1.239 & 4.206 & -5.000 & 2.247 & 20.000 & 1.221 & -0.143 \\
${}^3F_3$ & -100.000 & 1.624 & -2.361 & 4.206 & -10.000 & 2.247 &  &  &  \\
${}^3F_4$ & -60.000 & 1.720 & -2.219 & 4.000 & -10.000 & 2.400 &  &  &  \\
\hline
\end{tabular}
%\end{ruledtabular}
\end{table*}

\begin{table*} [!]
  \caption{Parameters of the neutron-$\He$ optical potential. \label{tab:2}
  }
%\begin{ruledtabular}
\setlength{\tabcolsep}{5pt}
%\begin{tabularx*}{\columnwidth}{l|*{9}{r}}
\begin{tabular}{l|*{9}{r}}
  \hline
  %                 1       2       9     10      7       8        5        6       13
  % arba            3       4      11     12      7*14                            
  ${}^{2S+1}L_J$ & $V_1$ & $R_1$ & $V_2$ & $R_2$ & $W_c$ & $R_w$ & $V_s$ & $R_s$  & $W_s$  
  \\ \hline
${}^1S_0$ &  -71.053 & 2.730 & 12.757 & 4.947 & -12.327 & 4.982 &  &  &  \\
${}^3P_0$ & -91.075 & 2.093 & 7.782 & 4.997 & -28.208 & 1.285 &  &  &  \\
${}^3S_1$ & 70.784 & 2.023 & 18.484 & 2.926 & -14.145 & 1.746 & 2.872 & 2.563 & -1.969 \\
${}^3D_1$ & -26.868 & 2.266 & 10.362 & 2.994 & -27.676 & 1.746 &  &  &  \\
${}^1P_1$ & -136.992 & 1.453 &        &      & -30.000 & 2.282 & 39.134 & 1.221 & -3.211 \\
${}^3P_1$ & -24.326 & 1.231 & -12.228 & 4.000 & -5.088 & 2.282 &  &  &  \\
${}^3P_2$ & -150.000 & 1.188 & -10.978 & 5.000 & -4.037 & 1.298 & -2.010 & 1.700 & 10.000 \\
${}^3F_2$ &          &       & -2.690 & 3.872 & -10.107 & 2.807  &  &  &  \\
${}^1D_2$ & 150.000 & 1.036 & -19.166 & 3.109 & -28.077 & 2.629 & 150.000 & 0.934 & -1.437 \\
${}^3D_2$ & -149.999 & 1.390 & 6.930 & 3.104 & -6.672 & 2.629 &  &  &  \\
${}^3D_3$ & -50.000 & 2.106 & 6.045 & 3.380 & -12.000 & 1.876 &  &  &  \\
${}^1F_3$ & -55.155 & 2.079 & 1.562 & 3.530 & -3.095 & 2.533 & 17.898 & 1.394 & -0.372 \\
${}^3F_3$ & -20.922 & 1.039 & -1.252 & 4.261 & -18.444 & 2.533 &  &  &  \\
${}^3F_4$ & -48.346 & 1.584 & -2.967 & 3.852 & -11.244 & 2.730 &  &  &  \\
\hline
\end{tabular}
%\end{ruledtabular}
\end{table*}

\begin{table*} [!]
  \caption{
Parameters of the proton-$\He$ optical potential in the presence of the Pauli term.
\label{tab:3}
  }
%\begin{ruledtabular}
\setlength{\tabcolsep}{5pt}
%\begin{tabularx*}{\columnwidth}{l|*{9}{r}}
\begin{tabular}{l|*{9}{r}}
  \hline
  %                 1       2       9     10      7       8        5        6       13
  % arba            3       4      11     12      7*14                            
  ${}^{2S+1}L_J$ & $V_1$ & $R_1$ & $V_2$ & $R_2$ & $W_c$ & $R_w$ & $V_s$ & $R_s$  & $W_s$  
  \\ \hline
${}^1S_0$ &  & & -17.639 & 3.544   & -4.732 & 0.700 &  &  &  \\
${}^3S_1$ & -96.946 & 1.165 & -4.142 & 3.691 & -0.768 & 2.344 & 4.929 & 2.617 & \
-0.919 \\
${}^3D_1$ &  &  & 3.980 & 3.691 & -8.060 &  2.344 &  &  &  \\
\hline
\end{tabular}
%\end{ruledtabular}
\end{table*}

\begin{table*} [!]
  \caption{
Parameters of the neutron-$\He$ optical potential in the presence of the Pauli term.
\label{tab:4}
  }
%\begin{ruledtabular}
\setlength{\tabcolsep}{5pt}
%\begin{tabularx*}{\columnwidth}{l|*{9}{r}}
\begin{tabular}{l|*{9}{r}}
  \hline
  %                 1       2       9     10      7       8        5        6       13
  % arba            3       4      11     12      7*14                            
  ${}^{2S+1}L_J$ & $V_1$ & $R_1$ & $V_2$ & $R_2$ & $W_c$ & $R_w$ & $V_s$ & $R_s$  & $W_s$  
  \\ \hline
${}^3S_1$ & -4.022 & 1.830 & -35.936 & 2.715 & -3.017 & 1.733 & 6.516 & 2.686 & \
1.552 \\
${}^3D_1$ & -64.913 & 1.830 & 21.735 & 2.715 & -19.226 &  1.733 &  &  &  \\
\hline
\end{tabular}
%\end{ruledtabular}
\end{table*}

%$J=11$  ${}^1S_0$ &  
%$J=12$  ${}^3P_0$ & 
%$J=13$  ${}^3S_1$ & 
%$J=13_2$  ${}^3D_1$ & 
%$J=14$  ${}^1P_1$ & 
%$J=14_2$  ${}^3P_1$ & 
%$J=15$  ${}^3P_2$ & 
%$J=15_2$  ${}^3F_2$ & 
%$J=16$  ${}^1D_2$ & 
%$J=16_2$  ${}^3D_2$ & 
%$J=17$  ${}^3D_3$ & 
%$J=18$  ${}^1F_3$ & 
%$J=18_2$  ${}^3F_3$ & 
%$J=19$  ${}^3F_4$ & 

%\clearpage
%%%%%%%%%%%%%%%%%%%%%%%%%%%%%%%%%%%%%%%%%%%%%%%%%%%%%%%%%%%%%%%%%%%%%%%%%%%%%

 \begin{figure}[h]
\begin{center}
\includegraphics[scale=0.64]{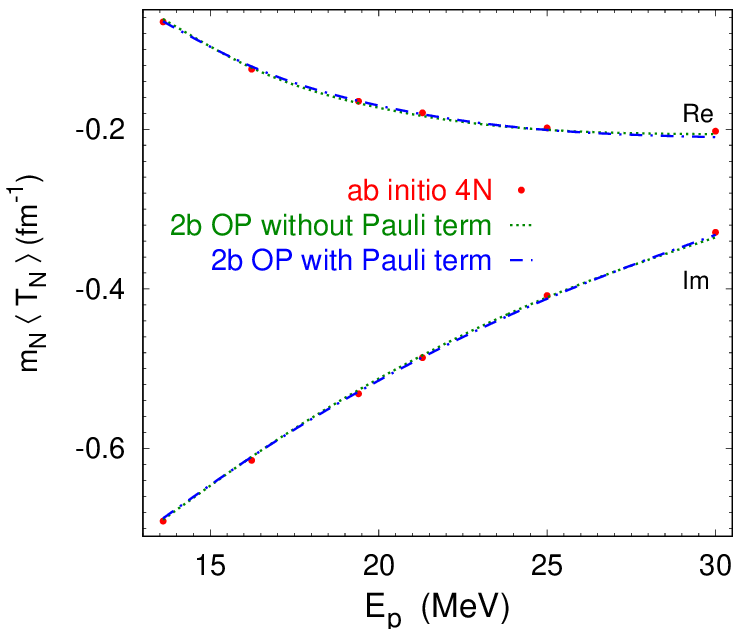}
\end{center}
\caption{\label{fig.ps}
Real and imaginary parts of the 
proton-$\He$ elastic scattering amplitude (after separating the point Coulomb contribution)
in the ${}^1S_0$ partial wave  as functions of the beam energy.
  Results of microscopic four-nucleon calculations (points) are
  compared with predictions obtained using two-body optical potentials (2b OP),
  either without (dotted curves) or with  (dashed-dotted curves) the Pauli term.
}
\end{figure}

 \begin{figure}[h]
\begin{center}
\includegraphics[scale=0.64]{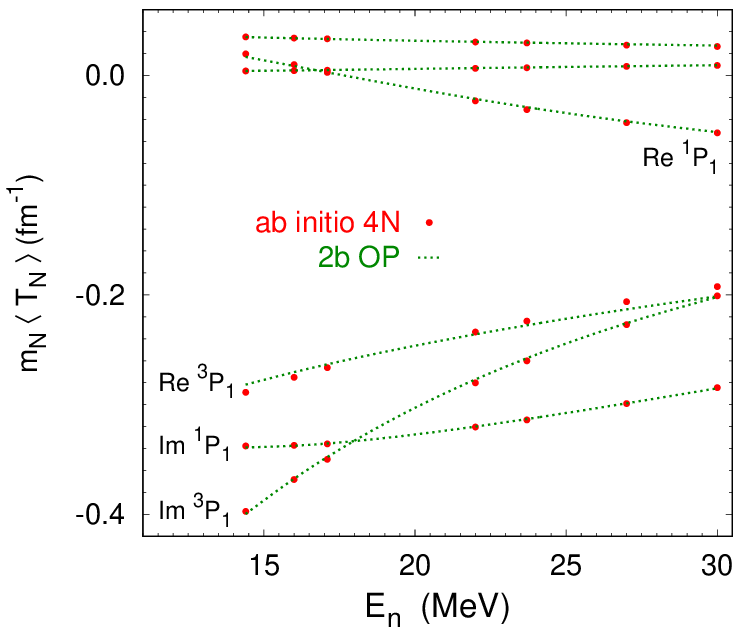}
\end{center}
\caption{\label{fig.ns}
Real and imaginary parts of the 
neutron-$\He$ elastic scattering amplitudes 
in the coupled ${}^1P_1 - {}^3P_1$ partial wave  as functions of the beam energy.
  Results of microscopic four-nucleon calculations (points) are
  compared with predictions obtained using two-body optical potentials (dotted curves).
The diagonal amplitudes are labelled, the non-diagonal ones are located near the zero line.
}
\end{figure}

\vspace{1mm}

%This work has received funding from the 
%Research Council of Lithuania (LMTLT) under Contract No.~S-MIP-22-72.

%\bibliographystyle{plbsty} 
%\bibliography{abbrev,pre80,80-89,90-99,200x,ad,book,clmb,exp,nreact,nstruct,4N,4Nex,dalpha} \end{document}